\documentclass[runningheads,a4paper]{llncs}

\usepackage{amssymb}
\usepackage{bm,bbm}
\usepackage{graphicx,subfigure}
\usepackage{booktabs}
\usepackage{multirow}
\usepackage[ugly]{units}
\usepackage{microtype}

\usepackage{url}
\urldef{\mails}\path|ljmtorres, medeiros.tonny, acfrery}@gmail.com|    
\newcommand{\keywords}[1]{\par\addvspace\baselineskip
\noindent\keywordname\enspace\ignorespaces#1}

\begin{document}

\title{Polarimetric SAR Image Smoothing with Stochastic Distances}

\titlerunning{Polarimetric SAR Image Smoothing with Stochastic Distances}

\author{Leonardo Torres\and Antonio C.\ Medeiros \and Alejandro C.\ Frery\thanks{The authors are grateful to CNPq and Fapeal for supporting this research}}

\authorrunning{Torres, Medeiros \& Frery}

\institute{
Universidade Federal de Alagoas -- UFAL\\
Laborat\'orio de Computa\c c\~ao Cient\'ifica e An\'alise Num\'erca -- LaCCAN\\
57072-970, Macei\'o, AL -- Brazil
}

\toctitle{Polarimetric SAR Image Smoothing with Stochastic Distances}
\tocauthor{Torres, Medeiros \& Frery}
\maketitle

\begin{abstract}
Polarimetric Synthetic Aperture Radar (PolSAR) images are establishing as an important source of information in remote sensing applications.
The most complete format this type of imaging produces consists of complex-valued Hermitian matrices in every image coordinate and, as such, their visualization is challenging. 
They also suffer from speckle noise which reduces the signal-to-noise ratio. 
Smoothing techniques have been proposed in the literature aiming at preserving different features and, analogously, projections from the cone of Hermitian positive matrices to different color representation spaces are used for enhancing certain characteristics. 
In this work we propose the use of stochastic distances between models that describe this type of data in a Nagao-Matsuyama-type of smoothing technique. 
The resulting images are shown to present good visualization properties (noise reduction with preservation of fine details) in all the considered visualization spaces.
\end{abstract}

\section{Introduction}\label{sec:intro}

Among the remote sensing technologies, PolSAR has achieved a prominent position. 
PolSAR imaging is a well-developed coherent and microwave remote sensing technique for providing large-scaled two-dimensional (2-D) high spatial resolution images of the Earth’s surface reflectivity; see Lee and Pottier~\cite{LeePottier2009Book}.

The phenomenon speckle in SAR data hinders the interpretation these data and reduces the accuracy of segmentation, classification and analyses of objects contained within the image.
Therefore, reducing the noise effect is an important task, and multilook processing is often used for this purpose in single-channel data.

According to Lee and Pottier~\cite{LeePottier2009Book}, the principle to preserve the polarimetric signature and Polarimetric SAR image smoothing requires:
(i)~for each element of the image should be filtered in a way similar to multilook processing by averaging the covariance matrix of neighboring pixels;
(ii)~the filtering should be executed independently for  each element of the covariance matrix; and
(iii)~homogeneous regions in the neighborhood should be adaptively selected to preserve resolution, edges and the image quality.

The statistical modeling provides a good support for the development of algorithms for interpreting PolSAR data efficiently, and for the simulation of plausible images.
Frery et al.~\cite{Frery2011HypothesisTest,Frery2011InformationTheoryPolSAR} introduce statistical tests for analyzing contrast in PolSAR images under the scaled multilook complex Wishart distribution, which has been successfully employed as a statistical model in such images for homogeneous regions.
Frery et al.~\cite{Frery2011InformationTheoryPolSAR} derive several distances and tests for the complex Wishart model.

This work presents a new smoothing process for PolSAR imagery based on stochastic distances and tests between distributions.
This process, beyond reducing the noise effect, maintains geometric features of the PolSAR data.
Vasile et al.~\cite{Vasile2006PolSAR} use a similar adaptive technique, but the decisions are based on the intensity information while we use the complete complex covariance matrix.

The paper is organized as follows:
In Section~\ref{sec:model} we summarise the model for polarimetric data.
Section~\ref{sec:distances} we describe the smoothing process for PolSAR images using stochastic distances between complex Wishart distributions, and the visualization of this kind of data.
Results are presented in Section~\ref{sec:results}, while Section~\ref{sec:conclu} concludes the paper.

\section{The Complex Wishart Distribution}\label{sec:model}

PolSAR imaging results in a complex scattering matrix, which includes intensity and relative phase data~\cite{Frery2011InformationTheoryPolSAR}.
Such matrices have possibly four distinct complex elements, namely $S_{VV}$, $S_{VH}$, $S_{HV}$, and $S_{HH}$, where $H$ and $V$ refer to the horizontal and vertical wave polarization states, respectively.
The complex signal backscattered from each resolution cell is characterized by the $p$-tuple scattering matrix vector $\bm{y}$, where $p=3$ for a reciprocal medium ($S_{VH}=S_{HV}$); see Ulaby and Elachi~\cite{Ulaby1990RadarPolarimetriy}. 

Thus, we have a scattering complex random vector
$
\bm{y}=[S_{VV},S_{VH},S_{HH}]^t,
$
where $[\cdot]^t$ indicates vector transposition.
In PolSAR data, the speckle might be modeled as a multiplicative independent zero-mean complex circular Gaussian process that modules the scene reflectivity~\cite{Ulaby1990RadarPolarimetriy,Touzi2004ReviewPolarimetry}, whose probability density function is
$$
f(y;\bm{\Sigma})=\frac{1}{\pi^3 \vert\bm{\Sigma}\vert}\exp\bigl\{-y^*\bm{\Sigma}^{-1}y\bigr\},
$$
where $\vert \cdot \vert$ is the determinant, the superscript `$*$' denotes the complex conjugate transpose of a vector, $\bm{\Sigma}$ is the covariance matrix of $\bm{y}$.
The covariance matrix $\bm{\Sigma}$, besides being Hermitian and positive definite, has all the information which characterizes the backscattering under analysis.

Multilook processing is intended to enhance the signal-to-noise ratio, thus, is calculated the averaged over $L$ ideally independent looks of the same scene.
This results in the sample covariance matrix $\bm{Z}$ given by
$
\bm{Z}=L^{-1}\sum_{\ell=1}^{L}\bm{y}_{\ell} \bm{y}_{\ell}^{*},
$
where $L$ is the number of looks $\bm{y}_{\ell}$, for $\ell = \{1, 2, \dots , L\}$, and the superscript `$*$' denotes the complex conjugate transposition.

According to Anfinsen et al.~\cite{EstimationEquivalentNumberLooksSAR}, $\bm{Z}$ follows a multilook scaled complex Wishart distribution, denoted by $\bm{Z}\thicksim\mathcal{W}(\bm{\Sigma},L)$.
Having $\bm{\Sigma}$ and $L$ as parameters, it is characterized by the following probability density function:
\begin{equation}
 f_{\bm{Z}}(\bm{Z}';\bm{\Sigma},L) = \frac{L^{3L}\vert \bm{Z}'\vert^{L-3}}{\vert\bm{\Sigma}\vert^L \Gamma_3(L)} \exp\bigl\{-L\ \mathrm{tr}\bigl(\bm{\Sigma}^{-1} \bm{Z}'\bigr)\bigr\},
\label{eq:denswishart}
\end{equation}
where $\Gamma_3(L)=\pi^3 \prod^2_{i=0} \Gamma(L-i)$, $\Gamma(\cdot)$ is the gamma function, $\mathrm{tr}(\cdot)$ is the trace operator, and the covariance matrix of $\bm{Z}$ is given by
$$
\bm{\Sigma} = E\{\bm{y}\bm{y}^*\} = \left[\begin{array}{ccc}
                                    E\{S_1 S_1^*\}\; & E\{S_1 S_2^*\}\; & E\{S_1 S_3^*\} \\
                                    E\{S_2 S_1^*\}\; & E\{S_2 S_2^*\}\; & E\{S_2 S_3^*\} \\
                                    E\{S_3 S_1^*\}\; & E\{S_3 S_2^*\}\; & E\{S_3 S_3^*\} \\
                                    \end{array} \right],
$$
where $E\{\cdot\}$ and the superscript `$*$' denote expectation and complex conjugation, respectively.

\section{Stochastic Distances Filter}\label{sec:distances}

\begin{figure}[hbt]
\centering
\includegraphics[width=.48\linewidth, trim= 10 10 10 20]{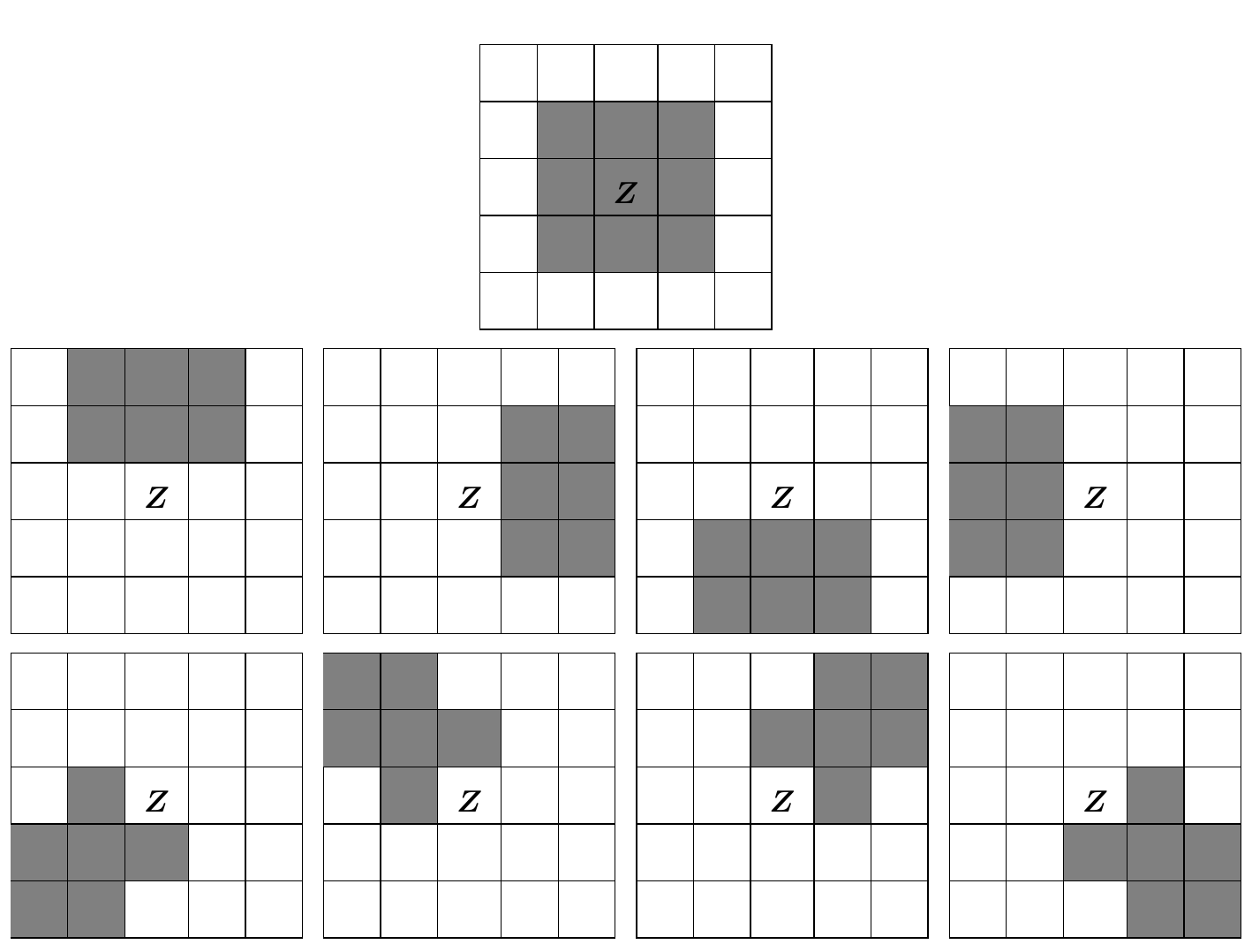}
\caption{Nagao-Matsuyama neighbourhoods.}
\label{fig:NagaoMatsuyama}
\end{figure}

The proposed filter is based on stochastic distances and tests between dis\-tri\-bu\-tions~\cite{Frery2011InformationTheoryPolSAR} obtained from the class of ($h,\phi$)-divergences.
It employs in a modified Nagao-Matsuyama set of neighbors~\cite{NagaoMatsuyama}, presented in Figure~\ref{fig:NagaoMatsuyama}.

The filtering procedure consists in checking which regions can be considered as coming from the same distribution that produced the data in the central block.
The sets which are not rejected are used to compute a local mean of covariance matrices.
If all the sets are rejected, the filtered value is updated with the average on the central neighborhood around the filtered pixel.

Each filtered pixel has a $5\times5$ neighborhood, within which nine overlapping areas are defined.
Denote $\bm{\widehat{\theta}_1}$ the estimated parameter in the central $3\times3$ neigh\-bor\-hood, and $\big(\bm{\widehat{\theta}}_2,\ldots,\bm{\widehat{\theta}}_{9}\big)$ the estimated parameters in the eight re\-main\-ing areas.

We estimate $\bm{\widehat{\theta}}_i=(\widehat{\bm{\Sigma}}_i)$ by maximum likelihood, assuming that the number looks $L$ is known; details can be seen in Anfinsen et al.~\cite{EstimationEquivalentNumberLooksSAR}.
Based on a random sample of size $n$, let $\{\bm{Z}_1,\bm{Z}_2,\dots,\bm{Z}_n\}$, the likelihood function related to the $\mathcal{W}(\bm{\Sigma},L)$ distribution is given by
\begin{equation}
  \mathcal{L}(\bm{\Sigma};\bm{Z}') = \Big( \frac{L^{3L}}{\vert\bm{\Sigma}\vert^L \Gamma_3(L)} \Big)^n \;\prod_{j=1}^{n} \vert \bm{Z}'\vert^{L-3} \exp\bigl\{-L\ \mathrm{tr}\bigl(\bm{\Sigma}^{-1} \bm{Z}'\bigr)\bigr\}.
\label{eq:LFwishart}
\end{equation}
Thus, the maximum likelihood estimator for $\bm{\Sigma}$ is $\widehat{\bm{\Sigma}}=n^{-1}\sum_{j=1}^{n}\bm{Z}_j$.

The proposal is based on the use of stochastic distances on small areas within the filtering window.
Consider that $\bm{Z}_1$ and $\bm{Z}_i$ are random matrices defined on the same probability space, whose distributions are characterized by the den\-si\-ties $f_{\bm{Z}_1}(\bm{Z}';\bm{\theta}_1)$ and $f_{\bm{Z}_i}(\bm{Z}';\bm{\theta}_i)$, res\-pec\-ti\-ve\-ly, where $\bm{\theta}_1$ and $\bm{\theta}_i$ are parameters.
Assuming that the densities have the same support given by the cone of Hermitian positive definite matrices $\bm{\mathcal{A}}$, the $h$-$\phi$ divergence between $f_{\bm{Z}_1}$ and $f_{\bm{Z}_i}$ is given by
\begin{equation}
D_{\phi}^{h}(\bm{Z}_1,\bm{Z}_i) = h \Big( \int_{\bm{\mathcal{A}}}\phi \Big( \frac{f_{\bm{Z}_1}(\bm{Z}';\bm{\theta}_1)}{f_{\bm{Z}_i}(\bm{Z}';\bm{\theta}_i)} \Big) \;f_{\bm{Z}_i}(\bm{Z}';\bm{\theta}_i)\;\mathrm{d}\bm{Z}' \Big),
\end{equation}
where $h\colon(0,\infty)\rightarrow[0,\infty)$ is a strictly increasing function with $h(0)=0$ and $h'(x)>0$ for every $x \in \mathbbm{R}$, and $\phi\colon (0,\infty)\rightarrow[0,\infty)$ is a convex function~\cite{Salicru1994}.
Choices of functions $h$ and $\phi$ result in several divergences.

Divergences sometimes are not dis\-tan\-ces because they are not symmetric.
A simple solution, described in~\cite{Frery2011HypothesisTest,Frery2011InformationTheoryPolSAR,Nascimento2010}, is to define a new measure $d_{\phi}^{h}$ given by
\begin{equation}
d_{\phi}^{h}(\bm{Z}_1,\bm{Z}_i) = \frac{D_{\phi}^{h}(\bm{Z}_1,\bm{Z}_i)+D_{\phi}^{h}(\bm{Z}_i,\bm{Z}_1)}{2}.
\end{equation}
Distances, in turn, can be conveniently scaled to present good statistical prop\-er\-ties that make them suitable as test statistics~\cite{Frery2011InformationTheoryPolSAR,Nascimento2010}:
\begin{equation}
\mathcal{S}_{\phi}^{h}(\bm{\widehat{\theta}}_1,\bm{\widehat{\theta}}_i)=\frac{2mnk}{m+n}\;d^{h}_{\phi}(\bm{\widehat{\theta}}_1,\bm{\widehat{\theta}}_i),
\end{equation}
where $\bm{\widehat{\theta}}_1$ and $\bm{\widehat{\theta}}_i$ are maximum likelihood estimators based on samples size $m$ and $n$, respectively, and $k=(h'(0)\phi'') ^{-1}$.
When $\bm\theta_1=\bm\theta_i$, under mild conditions $\mathcal{S}_{\phi}^{h}(\bm{\widehat{\theta}}_1,\bm{\widehat{\theta}}_i)$ is asymptotically $\chi^2_M$  distributed, being $M$ the dimension of $\bm{\theta}_1$.
Ob\-serv\-ing $\mathcal{S}_{\phi}^{h}(\bm{\widehat{\theta}}_1,\bm{\widehat{\theta}}_i)=s$, the null hypothesis $\bm{\theta}_1=\bm{\theta}_i$ can be rejected at level $\eta$ if $\Pr( \chi^2_{M}>s)\leq \eta$.
Details can be seen in the work by Salicr\'u et al.~\cite{Salicru1994}.

Since we are using the same sample for eight tests, we modified the value of $\eta$ by a Bonferroni-like correction, namely, the \u{S}id\'ak correction, that is given by $\eta = 1 - (1 - \alpha)^{1/t}$, where $t$ is the number of tests and, $\alpha$ the level of significance for the whole series of tests.

Frery et al.~\cite{Frery2011InformationTheoryPolSAR} derived several distances for the $\mathcal{W}(\bm{\Sigma},L)$ model, the one presented in Equation~(\ref{eq:denswishart}) among them.
The statistical test used in this paper was derived from the Hellinger distance, yielding:
\begin{equation}
 \mathcal{S}_{H} = \frac{8mn}{m+n}\Bigg[ 1-\bigg( \frac{\big\vert \big( \frac{\bm{\Sigma}_1^{-1} + \bm{\Sigma}_i^{-1}}{2} \big)^{-1} \big\vert}{\sqrt{\vert\bm{\Sigma}_1\vert\;\vert\bm{\Sigma}_i\vert}} \bigg)^L \Bigg].
\end{equation}

PolSAR is used to measure the target's reflectivity with four polarization channel combinations ($HH$, $HV$, $VH$ and $VV$), which can be expressed as a complex scattering matrix~\cite{LeePottier2009Book,Ulaby1990RadarPolarimetriy}.
Transformations on these channels polarization makes it possible to visualize the PolSAR data as a color image.

Two ways of visualizing the covariance matrix in false color are the Pauli (in the horizontal basis) and Sinclair decompositions.
They assign  $\vert S_{HH}-S_{VV} \vert^2 $, $\vert 2S_{HV} \vert^2$ and $\vert S_{HH}+S_{VV}\vert^2$, and $\vert S_{VV}\vert$, $\vert 2S_{HV}\vert$ and $\vert S_{HH}\vert$ to the red, green and blue channels, respectively.

\section{Results}\label{sec:results}

\begin{figure}[hbt]
\centering
  \subfigure[PolSAR data]{\label{fig:PauliOriginal}
  \includegraphics[width=.44\linewidth,viewport= 51 1 950 900,clip=TRUE]{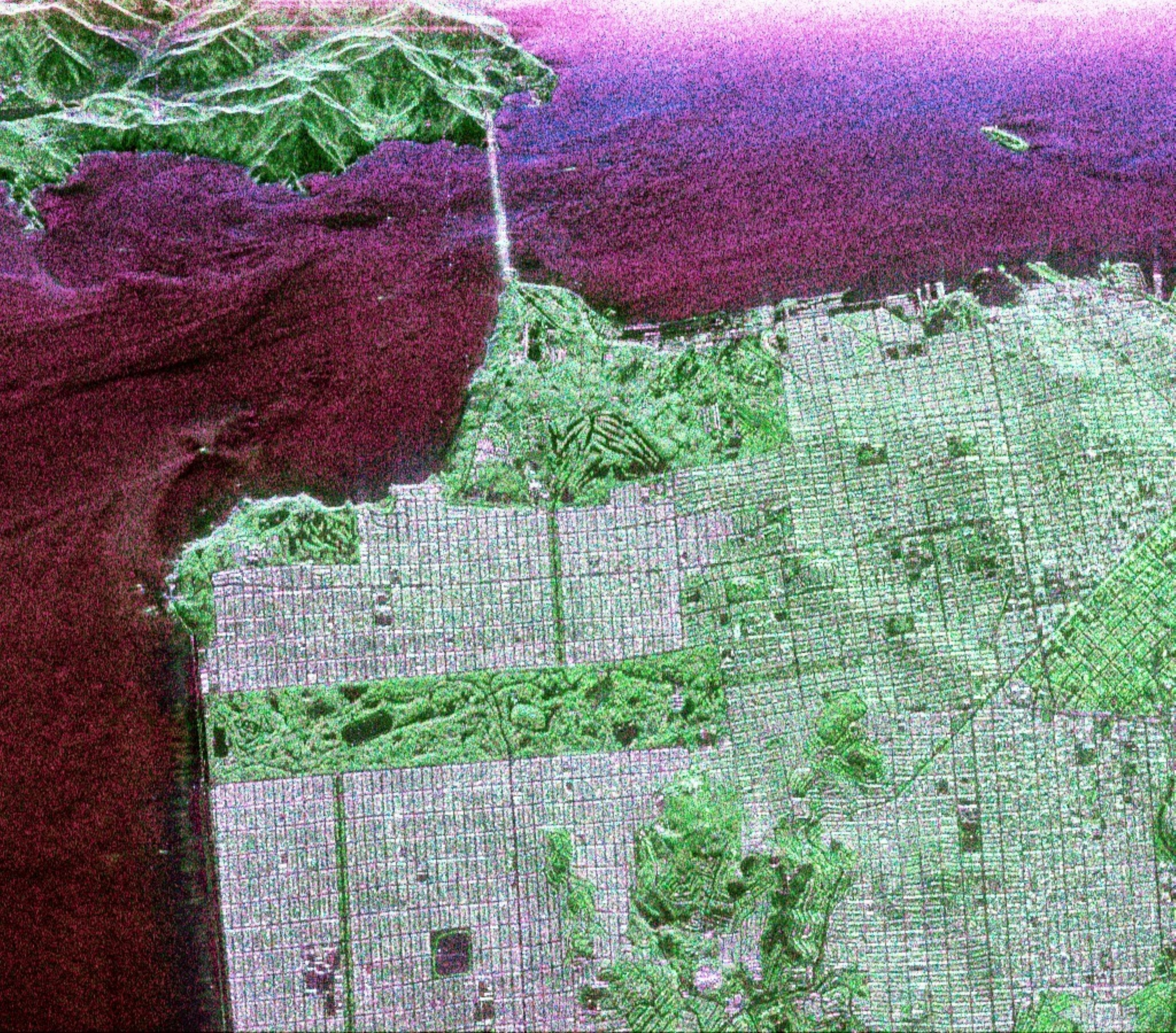}}%
  \subfigure[Zoom PolSAR data]{\label{fig:CutPauliOriginal}
  \includegraphics[width=.44\linewidth,viewport= 320 320 670 670,clip=TRUE]{figs/PauliOriginal.pdf}}

  \subfigure[Mean filter]{\label{fig:PauliFilterMean}
  \includegraphics[width=.44\linewidth,viewport= 51 1 950 900,clip=TRUE]{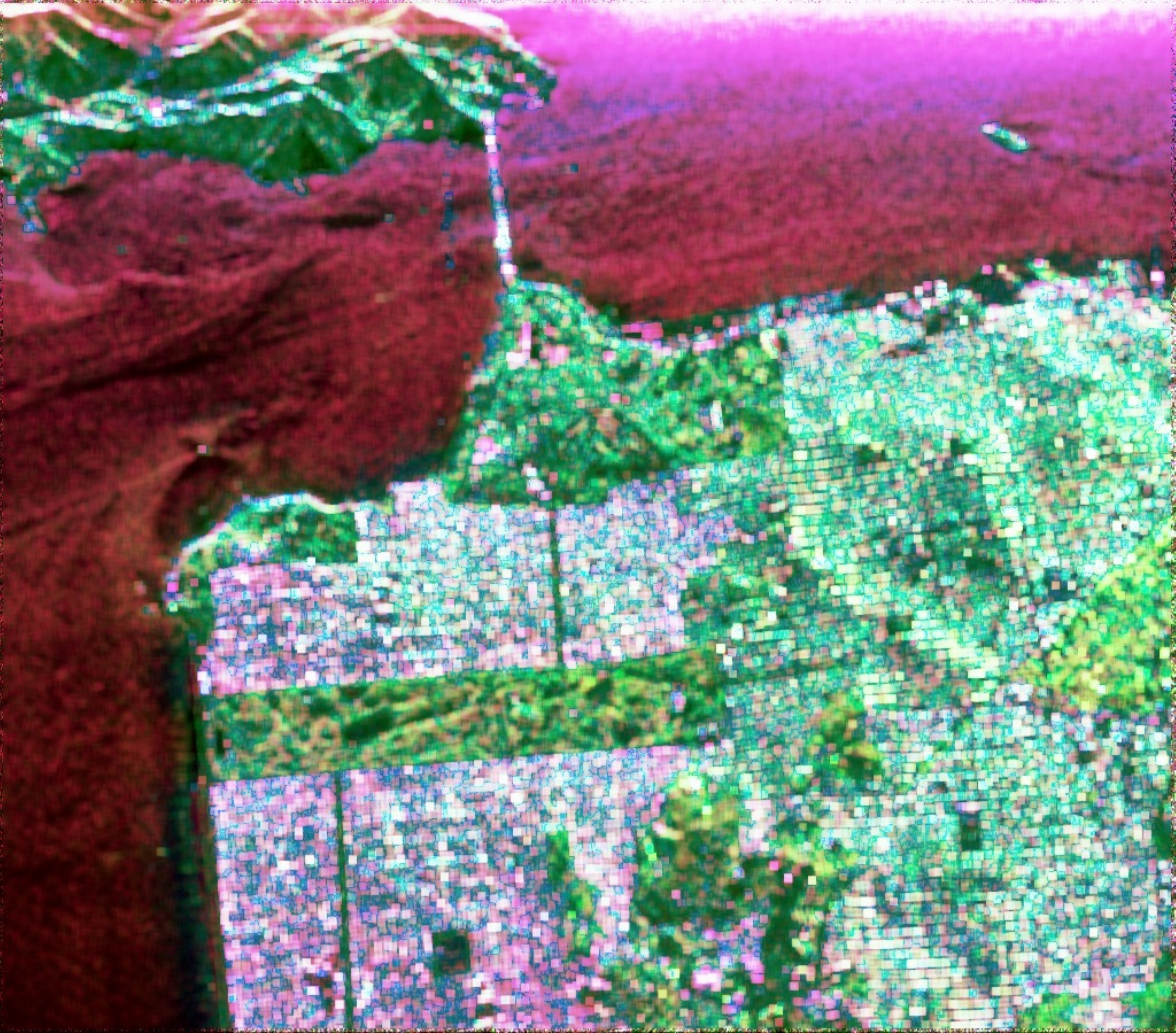}}%
  \subfigure[Zoom Mean filter]{\label{fig:CutPauliFilterMean}
  \includegraphics[width=.44\linewidth,viewport= 320 320 670 670,clip=TRUE]{figs/PauliFilterMean.pdf}}%

  \subfigure[Stochastic Distances filter]{\label{fig:PauliFilterH}
  \includegraphics[width=.44\linewidth,viewport= 51 1 950 900,clip=TRUE]{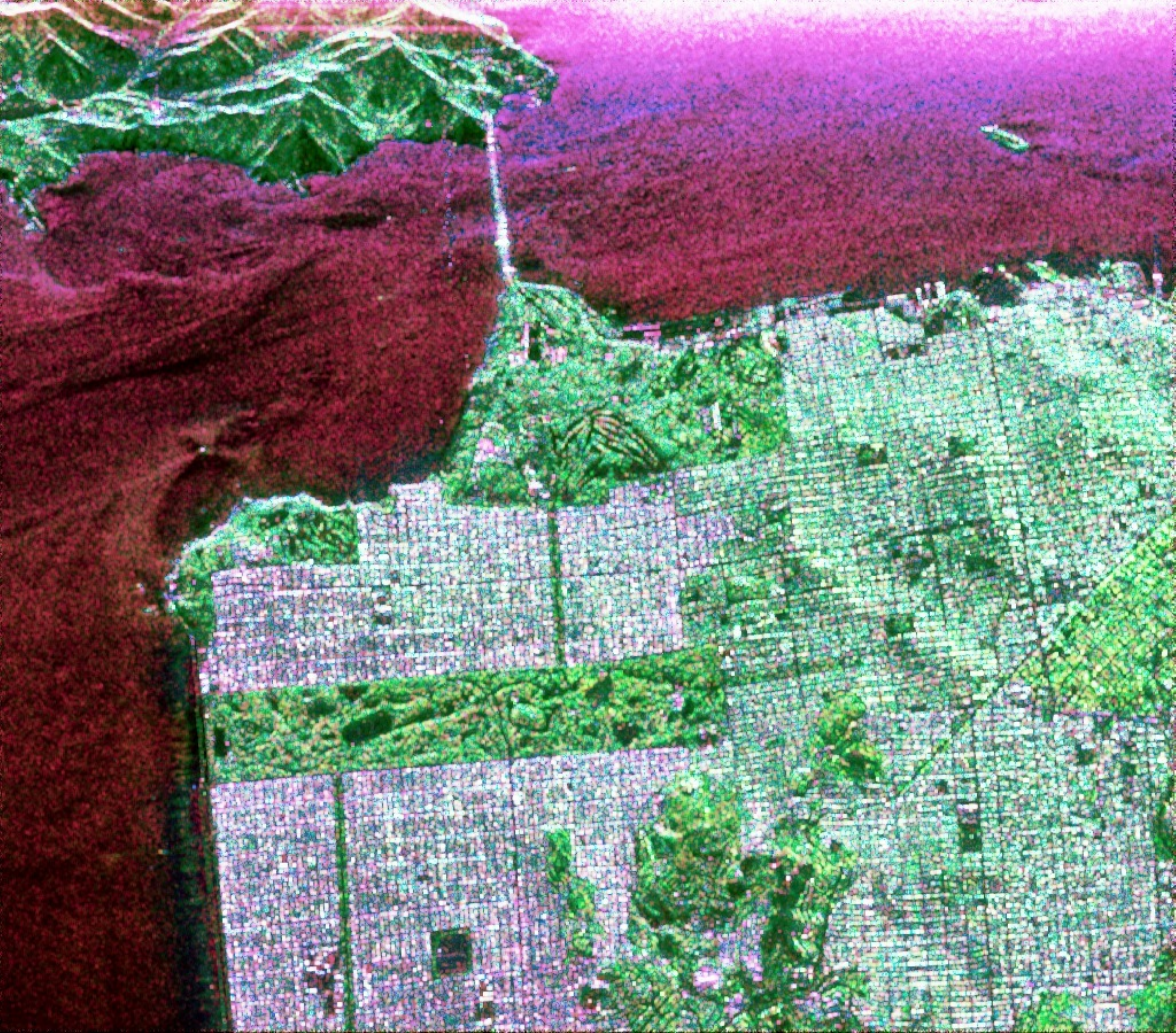}}%
  \subfigure[Zoom Stochastic Distances filter]{\label{fig:CutPauliFilterH}
  \includegraphics[width=.44\linewidth,viewport= 320 320 670 670,clip=TRUE]{figs/PauliFilterH.pdf}}%
\caption{PolSAR data on Pauli Decomposition.}
\label{fig:PolSARvisualizePauli}
\end{figure}

\begin{figure}[hbt]
\centering
  \subfigure[PolSAR data]{\label{fig:SinclairOriginal}
  \includegraphics[width=.44\linewidth,viewport= 51 1 950 900,clip=TRUE]{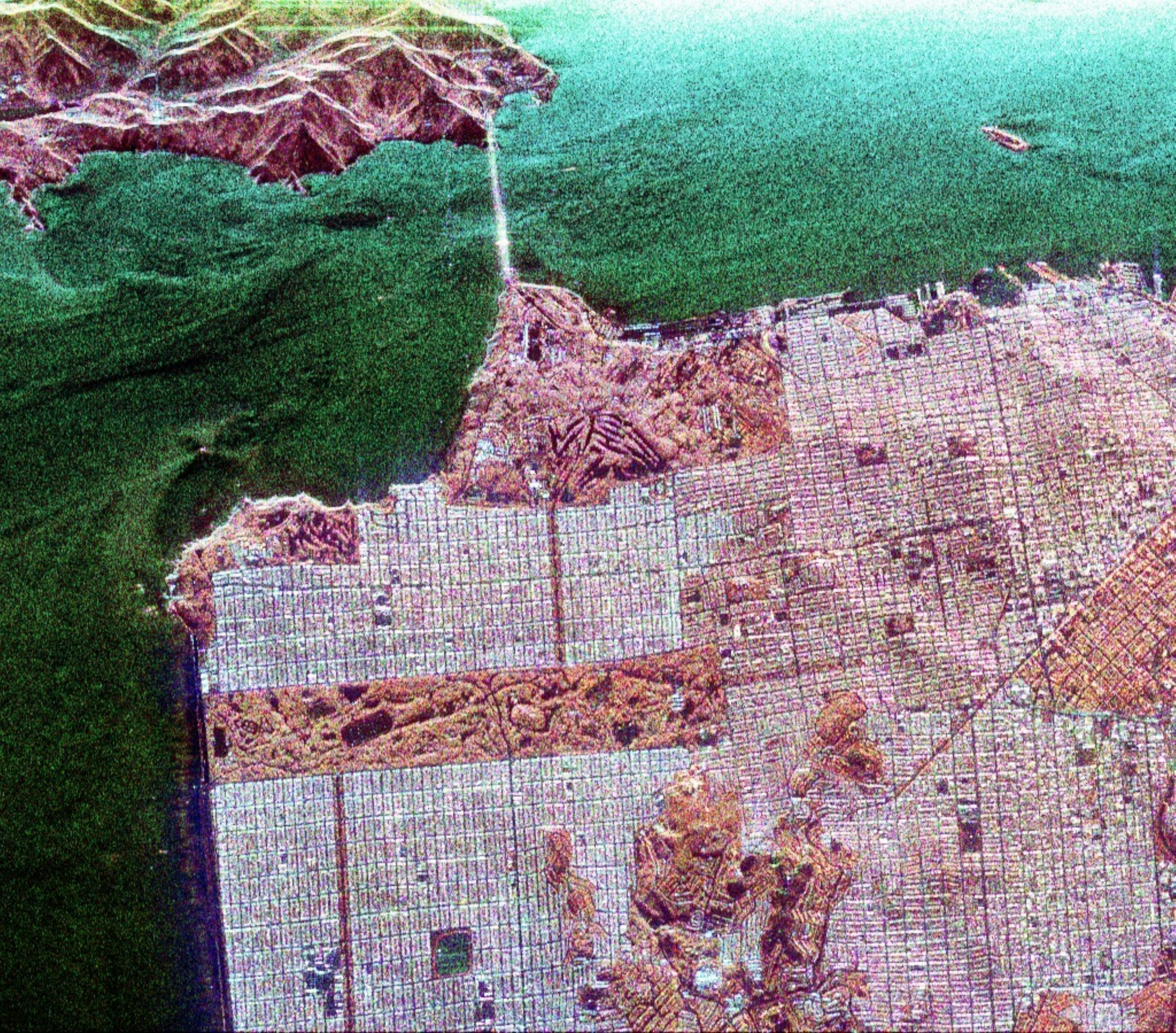}}%
  \subfigure[Zoom PolSAR data]{\label{fig:CutSinclairOriginal}
  \includegraphics[width=.44\linewidth,viewport= 320 320 670 670,clip=TRUE]{figs/SinclairOriginal.pdf}}

  \subfigure[Mean filter]{\label{fig:SinclairFilterMean}
  \includegraphics[width=.44\linewidth,viewport= 51 1 950 900,clip=TRUE]{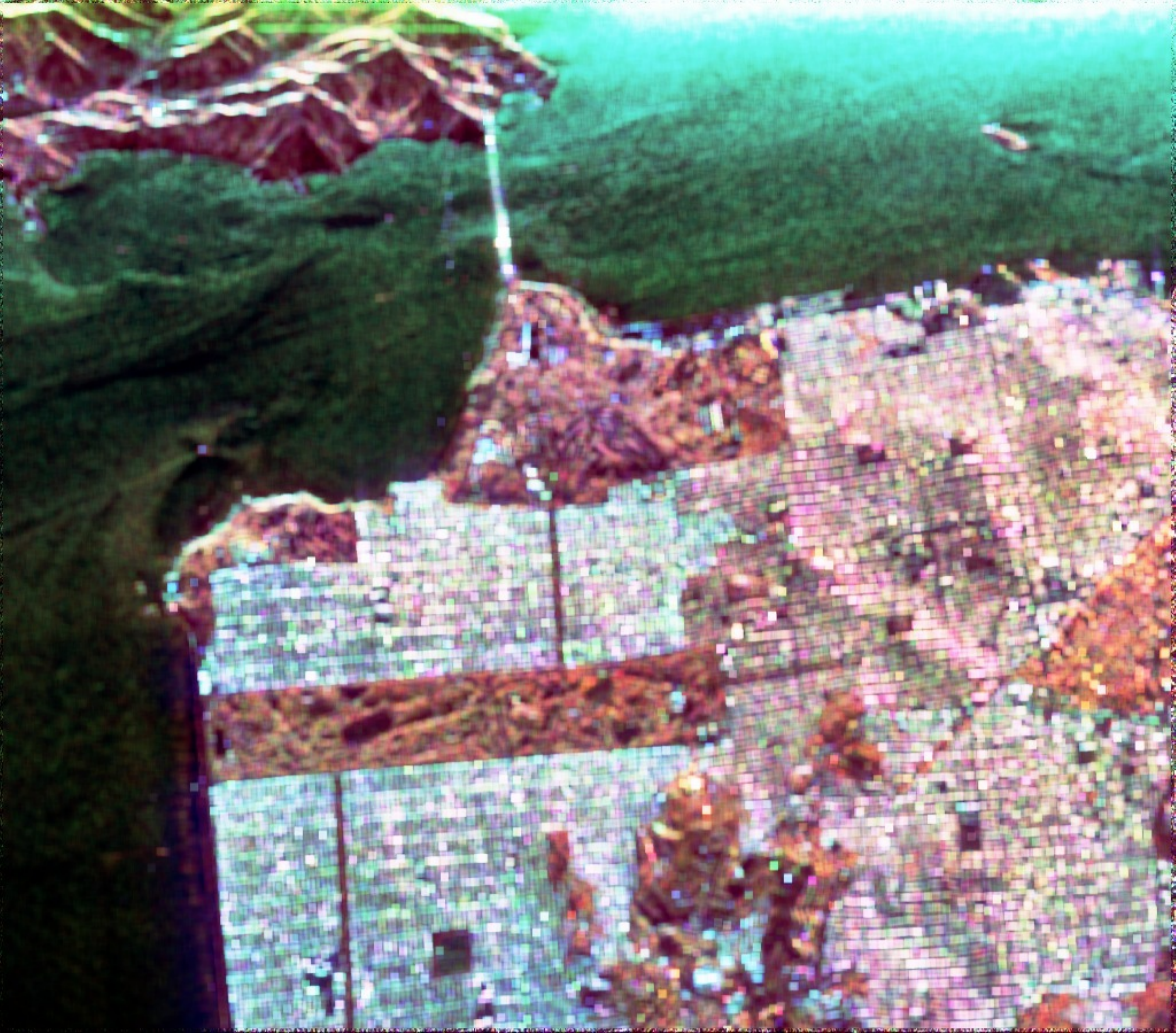}}%
  \subfigure[Zoom Mean filter]{\label{fig:CutSinclairFilterMean}
  \includegraphics[width=.44\linewidth,viewport= 320 320 670 670,clip=TRUE]{figs/SinclairFilterMean.pdf}}%

  \subfigure[Stochastic Distances filter]{\label{fig:SinclairFilterH}
  \includegraphics[width=.44\linewidth,viewport= 51 1 950 900,clip=TRUE]{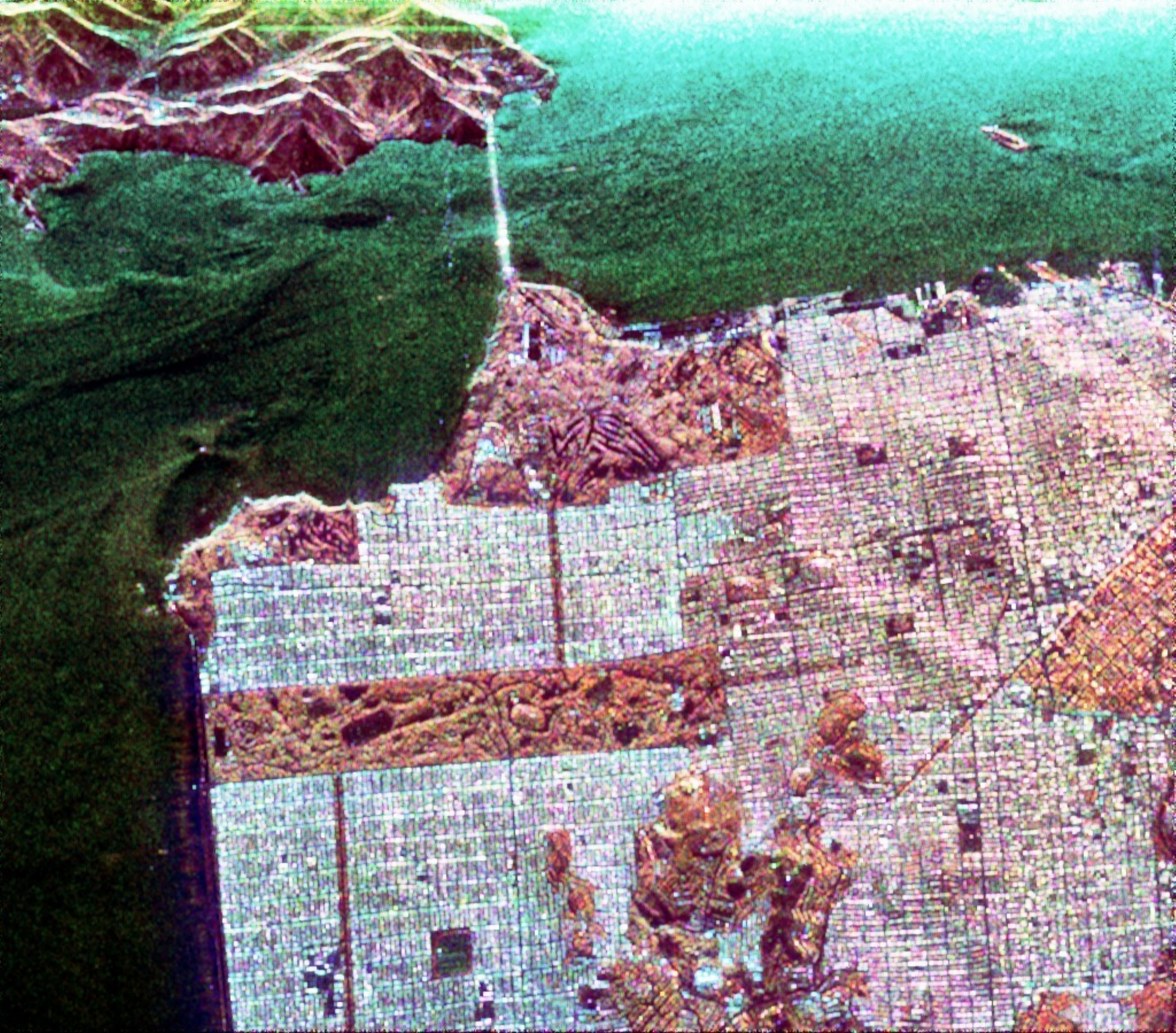}}%
  \subfigure[Zoom Stochastic Distances filter]{\label{fig:CutSinclairFilterH}
  \includegraphics[width=.44\linewidth,viewport= 320 320 670 670,clip=TRUE]{figs/SinclairFilterH.pdf}}%
\caption{PolSAR data on Sinclair Decomposition.}
\label{fig:PolSARvisualizeSinclair}
\end{figure}

The NASA/Jet Propulsion Laboratory Airborne SAR (AIRSAR) of the San Francisco Bay was used for evaluating the quality of the procedure.
The original polarimetric SAR data was generated with $4$-looks and $900\times1024$ pixels.
Figure~\ref{fig:PolSARvisualizePauli} presents results in the Pauli decomposition, while Figure~\ref{fig:PolSARvisualizeSinclair} shows their counterparts using the Sinclair decomposition.
Figures~\ref{fig:PauliOriginal} and~\ref{fig:CutPauliOriginal} (Figures~\ref{fig:SinclairOriginal} and~\ref{fig:CutSinclairOriginal}, respectively) show the original data set and a zoom.

Figures~\ref{fig:PauliFilterMean} and~\ref{fig:CutPauliFilterMean} (Figures~\ref{fig:SinclairFilterMean} and~\ref{fig:CutSinclairFilterMean}, resp.) show the effect of the mean computed on windows of size $5\times5$ over the whole image.
Albeit the noise reduction is evident, it is also clear that the blurring introduced eliminates useful information as, for instance, curvilinear details in the forested area.

Figures~\ref{fig:PauliFilterH} and~\ref{fig:CutPauliFilterH} (Figures~\ref{fig:SinclairFilterH} and~\ref{fig:CutSinclairFilterH}, resp.) present the result of smoothing the origintal data set computing means which pass the Hellinger test at the level significance $\alpha=80\%$.
The noise effect is alleviated, c.f.\ the reduced graininess specially in the forest and over the urban areas, but fine details are more preserved than when the mean is employed.
The directional selectiveness of the proposed filter retains linear structures as, for instance, the streets and the docks.
Bright structures within the forest are enhanced, and their linear appearance is maintained.

\section{Conclusions}\label{sec:conclu}

The proposed technique combats the effect of speckle noise in all the areas of the test image.
The filter is selective, producing stronger noise reduction in untextured areas, while it preserves fine details as linear structures and the forest texture. 
The proposal was compared with the simple mean filter using the decomposition process for PolSAR images.

Next steps will be assessing quantitatively the proposal, using iterated filters (since the complex Wishart distribution is closed under convolutions), and estimating the equivalent number of looks in order to consider possible departures from the homogeneous model.

% -------------------------------------------------------------------------
\bibliographystyle{splncs03}
\bibliography{ref_TorresCIARP2012,ref_PolSAR}

\end{document}